\documentclass[%
 reprint,
 superscriptaddress,
 amsmath,amssymb,
 aps,
 floatfix
 prl,
]{revtex4-2}

\usepackage{chemformula} 
\usepackage{xfrac}  
\usepackage{float} 
\usepackage{physics}

\usepackage{graphicx}
\usepackage{dcolumn}
\usepackage{bm}
\usepackage{xcolor}
\usepackage{soul}
\usepackage{CJKutf8}
\usepackage{verbatim}

\newcommand{\average}[1]{\langle{#1}\rangle}
\begin{document}
\begin{CJK}{UTF8}{gbsn}


\title{Understanding central spin decoherence due to interacting dissipative spin baths}

\author{Mykyta Onizhuk}
   \affiliation{Pritzker School of Molecular Engineering, University of Chicago, Chicago, IL 60637, USA}
\author{Yu-Xin Wang (王语馨)}
   \affiliation{Pritzker School of Molecular Engineering, University of Chicago, Chicago, IL 60637, USA}
   \affiliation{Joint Center for Quantum Information and Computer Science, University of Maryland, College Park, MD 20742, USA}
\author{Jonah Nagura}
   \affiliation{Pritzker School of Molecular Engineering, University of Chicago, Chicago, IL 60637, USA}
\author{Aashish A. Clerk}
   \affiliation{Pritzker School of Molecular Engineering, University of Chicago, Chicago, IL 60637, USA}

\author{Giulia Galli}%
   \email{gagalli@uchicago.edu.}
   \affiliation{Pritzker School of Molecular Engineering, University of Chicago, Chicago, IL 60637, USA}
   \affiliation{Department of Chemistry, University of Chicago, Chicago, IL 60637, USA}
   \affiliation{Materials Science Division and Center for Molecular Engineering, Argonne National Laboratory, Lemont, IL 60439, USA}

\date{\today}

\begin{abstract}
We propose a new approach to simulate the decoherence of a central spin coupled to an interacting dissipative spin bath with cluster-correlation expansion techniques. We benchmark the approach on generic 1D and 2D spin baths and find excellent agreement with numerically exact simulations. Our calculations show a complex interplay between dissipation and coherent spin exchange, leading to increased central spin coherence in the presence of fast dissipation. Finally, we model near-surface NV centers in diamond and show that accounting for bath dissipation is crucial to understanding their decoherence. Our method can be applied to a variety of systems and provides a powerful tool to investigate spin dynamics in dissipative environments.
\end{abstract}

\maketitle
\end{CJK}

\begin{figure*}
    \centering
    \includegraphics[scale=1]{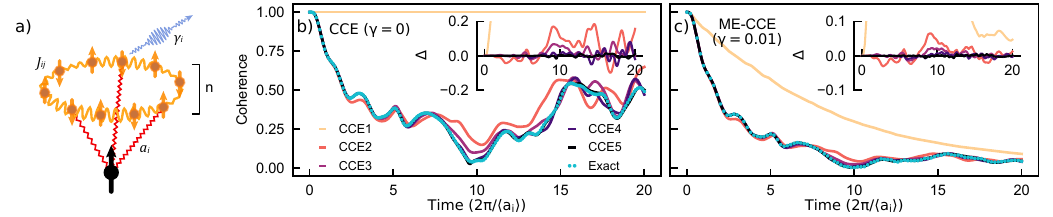}
    \caption{Decoherence of a central spin coupled to a dissipative spin chain. (a) Schematic of a twelve-spin chain ($n=12$) with uniformly distributed couplings  $J_{ij}\in[0,0.1 \cdot 2 \pi]$, $a_{i}\in[0,2 \cdot 2 \pi]$ (see text). 
    (b) Cluster-correlation expansion (CCE) simulations of central spin coherence coupled to a non-dissipative spin chain. CCE1 to CCE5 denote the order of the expansion. Blue points correspond to the full numerical simulations. (c) Master equation (ME)-CCE results of a central spin interacting with a dissipative spin chain (bath spin decay rate $\gamma=0.01$). Insets of (b) and (c) show the difference between the CCE predictions and exact numerical simulations $\Delta(t)=\mathcal{L}_{\text{CCE}}(t) - \mathcal{L}_{\text{exact}}(t)$ as a function of CCE orders (see text for the definition of $\mathcal{L}$).
}
    \label{fig:spinring}
\end{figure*}

Understanding the dynamical properties of open many-body quantum systems is a problem of fundamental importance in modern physics~\cite{Zoller_2008,Cirac_2009,Eisert_2015,Noel2022, Zu2021}. The interaction of a quantum system with an environment may radically change its behavior; for example, dissipation may drive novel phase transitions~\cite{Houck_2017,wu2023indication}, stabilize exotic states or subspaces~\cite{Zoller_2011,Schuster_2019,Hafezi_2019}, and lead to rich spectral statistics~\cite{PhysRevLett.123.254101}.
Recent experimental advances have enabled the ability to engineer dissipative environments \cite{Murch_2022}, as well as probe environmental effects via a single two-level system~\cite{Zu2021, Davis2023, PRXQuantum.3.040328}. Specifically, sensing experiments using single electron spins in solids have emerged as a hallmark near-term quantum technology, with applications in physics, chemistry, and biology~\cite{RevModPhys.89.035002, RevModPhys.92.015004}. 
 
Generic quantum sensing protocols seek to recover the dynamics of a many-body system from the dephasing of a two-level probe. Hence, to shed light on the main mechanisms determining the sensing signal, it is desirable to develop accurate computational methods to simulate the full dephasing dynamics generated by an interacting, many-body open quantum system. One may then use such computational techniques to solve a broad range of problems in materials science and quantum many-body physics.

In the case of closed quantum systems, the dynamical evolution of a two-level system (a central spin) coupled to a spin bath has been studied with the cluster-correlation expansion (CCE) technique \cite{PhysRevB.78.085315}. The CCE method is an efficient numerical approach that produces accurate results for many physical systems~\cite{Yang_2016, Seo2016, Kanai2022, PhysRevB.108.075306}. While alternate cluster techniques for translationally invariant systems have been used to study dissipative steady states \cite{PhysRevB.97.035103}, a more general method to investigate the dynamics of open quantum systems is not yet available.

In this work, we propose a novel approach for simulating dissipative, interacting spin baths and their impact on the central spin coherence evolution by solving many-body Lindblad master equations (ME) using CCE techniques. We benchmark the method against exact numerical simulations of baths formed by 1D and 2D lattices of spins with variable dissipation rates and also present results for realistic, experimentally motivated dissipative spin systems. 

\textit{Theoretical framework.}
We study the dynamics of a central spin coupled to an interacting spin bath with Markovian dissipation, as described by the GKSL (Lindblad) ME~\cite{Lindblad_1976,Gorini_1976} (in units $\hbar=1$):
\begin{equation}\label{eq:lindbladian}
    \frac{d}{dt} \hat \rho (t) = -i [\hat H, \hat \rho(t)] + \sum_i \gamma_i  \mathcal{D}[\hat L_i](\hat{\rho}),
\end{equation} 
where $\hat H = (\hat \sigma_z/2) \otimes \hat B + \hat H_{\text{bath}}$ is a pure dephasing interaction between the central spin and the bath operator $\hat B$,  $\hat H_\text{bath}$ describes intrabath interactions; $\mathcal{D}[\hat L_i](\hat{\rho})\equiv
\hat L_i \hat \rho \hat L_i^\dagger - \frac{1}{2}\{\hat L_i^\dagger \hat L_i, \hat \rho \}$ are superoperators accounting for incoherent processes between the bath and the external Markovian environment, and $\hat L_i$ are bath jump operators. Note that we do not assume the spin bath is translationally invariant or has a permutation symmetry.  
We omit any direct dephasing of the central spin (due to coupling to its own Markovian environment), as such dynamics can always be trivially factored out from the coherence.  


 The coherence function $\mathcal{L}(t)$ is defined as the normalized off-diagonal spin density matrix element $\mathcal{L}(t)=\Tr[\hat \rho_{01} (t)]/\Tr[\hat \rho_{01} (0)]$, where $\hat \rho_{01} (t) =\bra{0}\hat \rho (t)\ket{1}$ is a partial inner product of the total density matrix, and $\ket{0},\ket{1}$ are $\hat \sigma_z$ eigenstates of the central spin. Intuitively, $\mathcal{L}(t)$ represents the projection of the spin Bloch vector onto the $xy$-plane, $\mathcal{L}(t)=\average{\hat{\sigma}_{x}(t)}+i\average{\hat{\sigma}_y(t)}$ if the qubit is initially prepared in an equal superposition of basis states. Starting from Eq.~(\ref{eq:lindbladian}), the evolution of the $\hat \rho_{01} (t) $ is computed by solving~\cite{supplement}:
\begin{equation}\label{eq:mecoherence}
    \frac{d\hat \rho_{01} }{dt} = -i \hat H^{(0)} \hat \rho_{01} + i \hat \rho_{01} \hat H^{(1)} + \sum_i \gamma_i  \mathcal{D}[\hat L_i](\hat{\rho}_{01}),    
\end{equation} 
where $\hat H^{(\alpha)}=\bra{\alpha}\hat H \ket{\alpha}$ are effective Hamiltonians projected on the central spin states $\ket{\alpha}$.

Using the CCE method, one approximates the coherence function of the central spin interacting with the spin bath $\mathcal{L}(t)$ as a product of irreducible cluster contributions \cite{PhysRevB.78.085315, PhysRevB.79.115320}:
\begin{equation}\label{eq:l_cce}
	\mathcal{L}(t) = \prod_{C} \Tilde{\mathcal{L}}_C(t) = \prod_{i}\Tilde{\mathcal{L}}_{\{i\}}(t)\prod_{i,j}\Tilde{\mathcal{L}}_{\{ij\}}(t)...,
\end{equation}
where $\Tilde{\mathcal{L}}_{\{i\}}(t)$ is the contribution of a single bath spin~$i$, $\Tilde{\mathcal{L}}_{\{ij\}}(t)$ is the contribution of a spin pair~$i,j$ and so on. The largest size of the cluster defines the order of the approximation used. Each of the contributions is defined recursively from the coherence of the central spin, interacting with a single cluster $C$:
\begin{equation}\label{eq:l_contribution}
    \Tilde{\mathcal{L}}_C(t) = \frac{\mathcal{L}_{C}(t)}{\prod_{C'}\Tilde{\mathcal{L}}_{C'\subset C}(t)},
\end{equation}
where $C'$ indicates all subclusters of the cluster $C$. 
Conventionally, one computes the cluster coherence functions $\mathcal{L}_{C}(t)$ by solving the Schrodinger equation of the central spin coupled to a cluster~\cite{PhysRevB.78.085315}. 
Here, instead, we use Eq.~(\ref{eq:mecoherence}) to compute the cluster contributions in Eq.~(\ref{eq:l_contribution}) to account for the impact of both dissipative processes of individual bath spins and the coherent intrabath dynamics on the central spin coherence. We thus reduce the solution of a single intractable Lindbladian to a number of master equations of finite size equal to the order of the cluster expansion. We refer to such an approach as a ME-CCE.

\textit{Results.}
We start by benchmarking our method and first consider its general convergence properties. At short evolution times $t$ satisfying $\max _{\alpha} (||\hat H^{(\alpha)} | |) t \ll 1 $ and 
$\max _{i} (\gamma_i ||\hat L_i^\dagger \hat L_i | | ) t \ll1$, with $||\cdot||$ denoting the Frobenius operator norm, the ME-CCE method rapidly converges and the CCE may be truncated at a relatively low order, similar to conventional CCE calculations~\cite{supplement}. 
If the bath spin couplings can be grouped into disjoint clusters, the factorization and truncation in Eq.~\eqref{eq:l_cce} become exact as the order of the expansion increases. However, the order of truncation adopted in numerical simulations needs to be carefully checked in each specific case and depends on the parameter regime of interest.

We investigate the validity of our approach for the paradigmatic example of a central spin-$\frac{1}{2}$ coupled to a low-dimensional spin bath with local thermal baths and dipolar interactions.
We write the resulting XXZ-Hamiltonian under the secular approximation as:
\begin{equation}\label{eq:chain_ham}
    \hat H^{(0/1)} = \sum_{i=1}^{n} \pm \frac{a_i}{2} \hat I^i_z + \sum_{i,j}^{n} \frac{J_{ij}}{2} (\hat I^i_+ \hat I^j_- + \hat I^i_- \hat I^j_+ - 4\hat I^i_z \hat I^j_z ),
\end{equation}
where $\hat I^i$ are spin-$\frac{1}{2}$ operators of the $i$-th bath spin, $a_i$ are coupling of the $i$-bath spin to the central spin, $J_{ij}$ are dipolar couplings between nearest neighbours on the lattice.
The evolution of each cluster is computed with Eq.~(\ref{eq:mecoherence}), with $\hat L_{i,\pm}=\hat I^i_{\pm}$ as the bath spin jump operators. While our approach is fully general, for illustrative purposes, we consider the experimentally motivated regime where the bath is in the large-temperature limit. In this case, all bath spins thermalize into a maximally mixed state, and the raising and lowering spin jump operators have the same rate $\gamma_+=\gamma_-=\gamma$ (for examples outside of this regime see \cite{supplement}).

Consider first a one-dimensional spin chain as the bath (Fig.~\ref{fig:spinring}) in a parameter regime where the conventional CCE is expected to converge efficiently, i.e.~the couplings $a_i$ are significantly larger than the intrabath interactions $J_{ij}$~\cite{PhysRevB.78.085315}. 
The $a_i$ and $J_{ij}$ are chosen randomly, and the bath is initially quenched into a N\'eel state. 
We compare our results with the exact numerical simulation of the time evolution of a twelve-spin chain. If the spin bath is not dissipative, the CCE approach reproduces the exact dynamics at short timescales but shows deviation from the exact results at longer times, as expected. Increasing the order of the CCE approach reliably yields increasingly accurate results (Fig. \ref{fig:spinring}(b)).

We observe a similar behavior in the system with dissipative spins. The decay rate $\gamma$ is chosen to obtain sizeable contributions to the decoherence from both coherent and incoherent interactions in the bath (see Fig. \ref{fig:spinring}(c)). The ME-CCE coherence curve agrees well with the predictions of the direct numerical simulation results at higher orders of the approximation.

\begin{figure}
    \centering
    \includegraphics[scale=1]{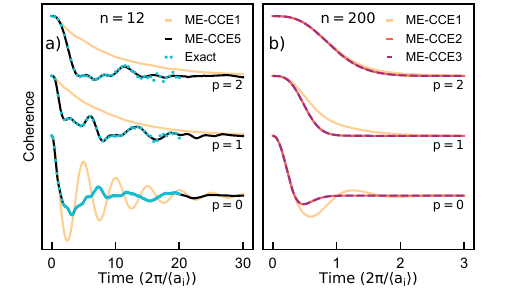}
    \caption{Spin decoherence due to a spin chain under dynamical decoupling. (a) Decoherence of a central spin coupled to a twelve-spin chain as a function of number of the $\pi$-pulses ($p$) applied to the central spin.
    (b) Decoherence of a central spin coupled to a large spin chain. The coupling parameters are chosen in the same fashion as in Fig. \ref{fig:spinring}(c).}
    \label{fig:largepn}
\end{figure}

We further test the predictions of our approach by simulating various pulse sequences applied to the central spin. Figure \ref{fig:largepn}(a) shows the comparison between the ME-CCE and the full numerical predictions for the Hahn-echo sequence ($p=1$), where a single $\pi$-pulse is applied to the central spin in the middle of the experiment, followed by results for two decoupling $\pi$-pulses ($p=2$). We note that sequences with a higher number of pulses require higher expansion orders to obtain converged results. 

The ME-CCE method allows us to study the decoherence of the central spin interacting with significantly larger chains, which would not be tractable with exact methods. Figure \ref{fig:largepn}(b) shows the decoherence of the central spin coupled to a chain with 200 spins ($n=200$) under different pulse sequences. Similar to the conventional CCE, the order required for convergence decreases as the size of the bath increases: the difference between truncating at the second or third order is negligible for all sequences studied here.

Finally, we find that the ME-CCE approach accurately captures the coherence dynamics even when collective dissipation is included in the simulation. For instance, we compute the dynamics of a chain with incoherent spin exchange between nearest neighbors,
    $\hat L_{i,\pm}=\hat I^i_\mp\otimes \hat I^{i+1}_\pm $
and we find that the ME-CCE results match the predictions of exact numerical simulations (see \cite{supplement}).


\begin{figure}
    \centering
    \includegraphics[scale=1]{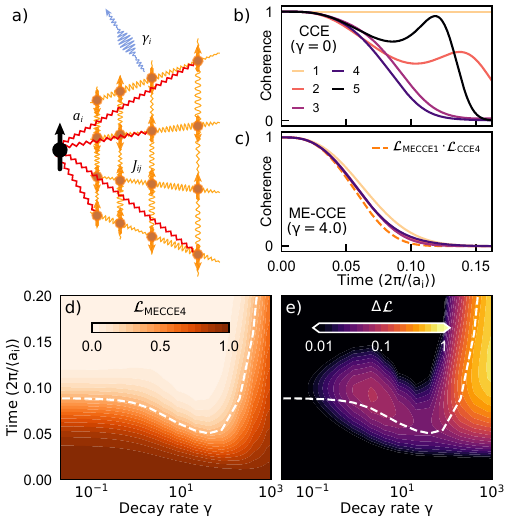}
    \caption{Decoherence of a central spin coupled to a dissipative 2D lattice of spins. (a) Schematic representation of the system under study where $J_{ij}$, $a_i$, and $\gamma_i$ are defined as in Fig.~\ref{fig:spinring}. (b) Hahn-echo signal with no dissipation. Different colors correspond to different cluster-correlation expansion (CCE) orders. (c) Hahn-echo signal of the central spin as a function of the CCE order. (d) Hahn-echo coherence as a function of the decay rate of the bath spins. (e) Difference between full and factorized coherence function (see text) in the Hahn-echo experiment as a function of the bath-spin decay rate $\gamma$. For all plots,
    the spin bath has 400 spins ($n=400$) and starts from an initial random product state.  Couplings are chosen as $J_{ij}=4 \cdot 2 \pi$, $a_{i}\in[0,2 \cdot 2 \pi]$.}
    \label{fig:torus}
\end{figure}

As a next example, we investigate the decoherence of the central spin interacting with a two-dimensional dipolar-coupled square spin lattice (Fig. \ref{fig:torus}(a)). 
In sensing applications, the electron spins commonly constitute the dominating spin bath at the surface of the host material~\cite{Grinolds2014}. Thus, it is of interest to understand whether the proposed approach is suitable to simulate the regime where the strength of the coupling between the central spin and the bath is comparable with that of intrabath interactions.

Such systems with no dissipative dynamics are challenging to simulate with the conventional CCE, requiring complex averaging procedures to reconstruct the coherence curve~\cite{PhysRevB.86.035452}. In contrast, we find that the presence of dissipative processes suppresses higher-order correlations, significantly improving the convergence of the ME-CCE method compared to that of coherent spin bath calculations (Fig. \ref{fig:torus}(b,c)). Hence, the ME-CCE approach can be used to study spin baths with a wide range of intrabath coupling strengths.

Figure \ref{fig:torus}(d) shows the Hahn-echo signal of the central spin as a function of the bath-spin decay rate $\gamma$. To characterize the decoherence process, we use the coherence time $T_2$, defined as the time at which the coherence function decays to $1/{e}$ of the initial value. We find that the coherence time varies nonmonotonically with $\gamma$: at low rates, it decreases with increasing $\gamma$, reaches a minimum near the coupling rate ($\gamma\sim a_i)$, and increases again at larger $\gamma$.
The $T_2$ enhancement in the large decay rate regime arises from the motional narrowing of the bath \cite{PhysRevB.72.045330}; these results for $T_2$ provide a microscopic model explaining the coherence improvement observed experimentally in the case of a stochastic drive of the spin bath \cite{Joos2022}. In this regime, the single spin contributions $\mathcal{L}_i$ completely dominate the decoherence of the central spin. We can integrate Eq. (\ref{eq:mecoherence}) for a single bath spin, assuming the bath is in the infinite temperature limit, and we write $\mathcal{L}_i$ explicitly during the free evolution of the central spin as (see, e.g.,~\cite{Jurcevic_2022}):
\begin{equation}\label{eq:analli}
    \mathcal{L}_i=e^{-\gamma_i t}\left( \cosh[\frac{1}{2}\Tilde\omega_i t] + \frac{2\gamma_i}{\Tilde\omega_i}\sinh[\frac{1}{2}\Tilde\omega_it]\right),
\end{equation}
where $\Tilde\omega_i = \sqrt{4\gamma_i^2 - a_i^2}$. In the small $\gamma_i$ limit, the central spin coherence decays exponentially with a decay rate equal to $\gamma_i$, and in the large $\gamma_i$ limit, the coherence decay rate is proportional to $a_i^2/\gamma_i$.
At decay rates larger than the half coupling $\gamma_i > \frac{a_i}{2}$, the r.h.s.~of Eq.~(\ref{eq:analli}) is strictly real, and the dephasing time of the central spin \textit{decreases} with increased $\gamma_i$, leading to an increase in the coherence time.

To investigate the interplay between single spin relaxation and coherent spin exchange over a wide range of parameters, we compare the results of the full ME-CCE simulation at fourth order $\mathcal{L}_{\text{MECCE4}}$ to the result obtained assuming two independent contributions. We compute the latter as a product of the contribution of the incoherent spin jumps from non-interacting spins, obtained with ME-CCE1, and the coherent spin exchange, computed with the conventional CCE at fourth order ($\mathcal{L}_{\text{MECCE1}} \cdot \mathcal{L}_{\text{CCE4}}$ in Fig.~\ref{fig:torus}(c)). The difference
$\Delta \mathcal{L} = \mathcal{L}_{\text{MECCE4}} - \mathcal{L}_{\text{MECCE1}} \cdot \mathcal{L}_{\text{CCE4}}$
allows us to quantify the interplay between the two processes.

\begin{figure}
    \centering
    \includegraphics[scale=1]{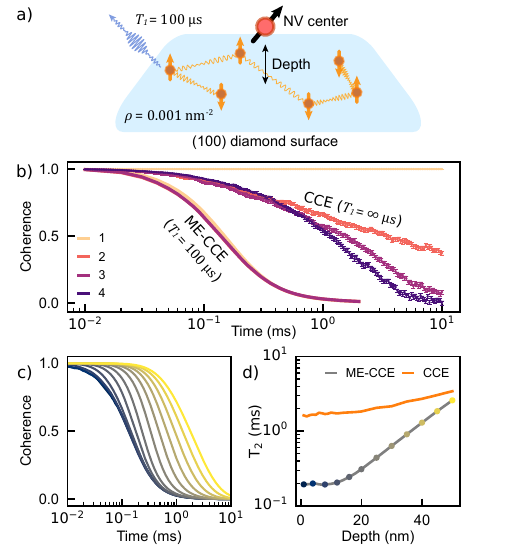}
    \caption{Decoherence of a near-surface NV center in diamond. (a) Schematic representation of the system ($\rho$ is the surface spin density, $T_1$ is the relaxation time of the surface spins). (b) The Hahn-echo signal computed at various orders of the cluster-correlation expansion (CCE, shown in color) of the NV center at a distance of 10 nm from the surface. Solid lines show results with the dissipation of bath spins, and points show the predictions of the conventional CCE method. (c) Coherence of the NV center as a function of the distance from the surface (distance varies from 0 to 50 nm in going from darker to lighted lines). (d) NV center coherence time $T_2$, obtained with the ME-CCE and CCE methods as a function of the distance from the surface.}
    \label{fig:diamond}
\end{figure}

Figure \ref{fig:torus}(e) shows the difference in coherence curves $\Delta \mathcal{L}$ as a function of the decay rate $\gamma$. 
We find that $\Delta \mathcal{L}$ is strictly positive, i.e., the factorized coherence is smaller than or equal to the complete coherence function, indicating that an approximate treatment always predicts a faster coherence decay. The presence of dissipative dynamics thus effectively suppresses the coherent interactions in the bath and mitigates the effect of coherent processes on the central spin decoherence dynamics.

We note that only when the strength of the central-spin-bath interactions is comparable to both the strength of exchange interactions and that of the bath dissipation do we observe a strong deviation from zero of  $\Delta \mathcal{L}$. In the regime where the couplings $a_i$ are significantly larger than the exchange terms in the bath $J_{ij}$, the effects of the coherent interactions and incoherent spin flips on the central spin decoherence are almost entirely separable~\cite{supplement}.


Finally, we apply our method to systems with practical relevance to sensing applications. We study the decoherence of the negatively charged nitrogen-vacancy (NV) center in diamond near the (100) surface (Fig. \ref{fig:diamond}(a)). The NV center is arguably the most studied and well-understood spin sensor, where room-temperature quantum manipulations have been routinely demonstrated \cite{RevModPhys.92.015004, Schirhagl2014}, and the (100) surface is the most common facet used for sensing applications
\cite{Maletinsky2012, PhysRevX.10.011003,PRXQuantum.3.040328, PhysRevApplied.20.014040}. We assume that the surface contains 0.001 nm$^{-2}$ electronic spins with the same gyromagnetic ratio and that each surface spin has an intrinsic relaxation to a fully depolarized state with a lifetime of $T_1=100$ $\mu$s. This set of parameters is chosen to match the conditions reported in Ref.~\cite{PRXQuantum.3.040328} and is consistent with those of other experimental measurements \cite{PhysRevLett.113.197601, PhysRevApplied.10.064045, PhysRevApplied.19.L031004, rezai2023probing}.

Under these conditions, we observe that the incoherent dynamics almost completely dominates the surface-induced decoherence of the NV center in the Hahn-echo experiment (Fig. \ref{fig:diamond}(b)). We also find a different qualitative behavior of the coherence time, obtained with and without dissipative dynamics, as a function of the distance from the surface (Fig. \ref{fig:diamond}(d)). This result highlights the need to include dissipative dynamics of bath spins for all future first-principles studies of surface-induced decoherence.

\textit{Discussion.}
In conclusion, we have introduced a cluster-correlation expansion approach to solve the Lindblad master equation and compute the coherence of the central spin in a dissipative spin bath. Our approach reproduces the results of full numerical simulations for several 1D and 2D models and shows a remarkably robust convergence in a wide range of coupling parameters. 

Our calculations revealed a rich interplay between Hamiltonian and dissipative dynamics in the bath and a nontrivial dependence of the central spin decoherence on the relaxation rate in the lattice. In addition, they provided a  microscopic model to explain the coherence improvements observed experimentally under stochastic driving of surface spins \cite{Joos2022}.

We also investigated a shallow NV center near the (100) diamond surface and found that the surface spins' incoherent relaxation completely dominates the NV center's spin-echo decoherence in the experimentally observed range of parameters. This finding highlights the importance of including dissipative dynamical processes in first-principles simulations and motivates further in-depth studies. 

We emphasize that the methodology presented here is general and applicable to any many-body open system coupled to a single probe. Areas of interest include probing the spin squeezing with a single two-level system~\cite{block2023universal} in a dissipative environment, the interplay between phonon-limited relaxation and coherent spin exchange in bulk materials at high temperature, as well as understanding the impact of incoherent spin hopping at surfaces.

\textit{Codes used.}
Full numerical simulations were carried out with QuTiP \cite{Johansson2013}. The ME-CCE approach will be made available as a part of the PyCCE package \cite{Onizhuk2021b}.


\textit{Acknowledgements}
We thank Jonathan Marcks for useful discussions. M.O., J.N., and G.G. acknowledge the support of NSF QuBBE Quantum Leap Challenge Institute (Grant No. NSF OMA-2121044).  We also acknowledge partial support by the University of Chicago Materials Research Science and Engineering Center, which is funded by the National Science Foundation under Grant No. DMR-2011854.
Code development in this work was supported by MICCoM, as part of the Computational Materials Sciences Program funded by the U.S. Department of Energy, Office of Science, Basic Energy Sciences, Materials Sciences, and Engineering Division through Argonne National Laboratory.

\bibliography{references}

\end{document}